\documentclass[journal,twocolumn]{IEEEtran}
 \usepackage[pdftex]{graphicx}

\ifCLASSINFOpdf
\else
\fi
\DeclareUnicodeCharacter{2212}{-}
\DeclareUnicodeCharacter{1D4C1}{-}
\DeclareUnicodeCharacter{FFFD}{-}

\begin{document}
%
\title{Low-rank Convex/Sparse Thermal Matrix Approximation for Infrared-based Diagnostic System}
\author{\IEEEauthorblockA{Bardia Yousefi $^\dagger$, 
Clemente Ibarra Castanedo, and 
Xavier P.V. Maldague}\\

\IEEEauthorblockA{Computer Vision and Systems Laboratory (CVSL), 
Department of Electrical and Computer Engineering, \\
Laval University, Quebec City (Quebec) G1V 0A6, Canada}\\

\IEEEauthorblockA{\textit{$^\dagger$ Address: }Department of Radiology, University of Pennsylvania, Philadelphia PA 19104}\\
\thanks{
This research was conducted by support of the Tier-1 Canadian research chair in Multipolar Infrared Vision (MIVIM), at Laval University.

B. Yousefi is with the Department of Electrical and Computer Engineering, Computer Vision and Systems Laboratory (CVSL), Laval University. His current address is at the Department of Radiology, University of Pennsylvania, Philadelphia PA 19104.
Corresponding author: B. Yousefi (e-mail: Bardia.Yousefi.1@ulaval.ca).

C. I. Castanedo is with the Department of Electrical and Computer
Engineering, Computer Vision and Systems Laboratory, Laval University (e-mail: IbarraC@gel.ulaval.ca).

X. P. V. Maldague is with the Department of Electrical and Computer
Engineering, Computer Vision and Systems Laboratory, Laval University (e-mail: Xavier.Maldague@gel.ulaval.ca).

This is the authors version of the article published in IEEE Transactions on Instrumentation and Measurement, doi: 10.1109/TIM.2020.3031129.}}

%



\IEEEtitleabstractindextext{%
\begin{abstract}
Non-negative matrix factorization (NMF) controls negative bases in the principal component analysis (PCA) with non-negative constraints for basis and coefficient matrices. Semi-, convex-, and sparse-NMF modify these constraints to establish distinct properties for various applications in different fields, particularly in infrared thermography. 
In this study, we delve into the applications of semi-, convex-, and sparse-NMF in infrared diagnostic imaging systems. We applied these approaches to active and passive thermographic imaging systems to determine the heterogeneous thermal patterns in these sets. In active thermography, three diverse specimens, carbon fiber-reinforced polymer composites (CFRP), poly(methyl methacrylate) (PMMA, also known as Plexiglas), and aluminum plate, were used. Quantitative analyses were performed using the Jaccard index. In passive thermography, 55 participants for infrared breast screening selected from the Database for Mastology Research (DMR) dataset with symptomatic and healthy participants. We calculated five derived properties of the breast area (contrast, correlation, dissimilarity, homogeneous, and energy) by using thermal level co-occurrence matrices (TLCMs) and trained a logistic regression method to stratify between healthy and symptomatic patients.
For both scenarios, we compared the ability of semi-, convex-, and sparse-NMF to state-of-the-art thermographic approaches, such as principal component analysis/thermography (PCT), candid covariance-free incremental principal component thermography (CCIPCT),  sparse- PCT, and NMF. Measurement of different defect depths and sizes indicated significant performance for sparse-NMF  (AL($d <1 mm$): 90.6$\%$, CFRP ($s>10 mm$): 42.7$\%$, PLEXI ($1 mm< d \leq 3 mm$): 46.9$\%$, DMR: 74.1$\%$), Semi-NMF (AL ($d \leq 1 mm$): 86.4$\%$, CFRP($s>1 mm$): 81.7$\%$, PLEXI ($1 mm < d \leq 3 mm$): 37.6$\%$, DMR: 75.8$\%$), and Convex-NMF (AL($d \leq 1 mm$): 91.2$\%$,CFRP ($s>10 mm$): 97.4$\%$, PLEXI ($d\leq1 mm$): 86.8$\%$, DMR: 77.8$\%$).  Moreover, we tested the robustness of all these algorithms against additive Gaussian noise (3$\%$ to 20$\%$) through the signal-to-noise-ratio (SNR). The results revealed considerable performance for semi-, convex- and sparse- NMF. Hence, the three methods exhibit promising performance in terms of accuracy and robustness for confirmation of the outlined properties.

\end{abstract}

}

\maketitle

\IEEEdisplaynontitleabstractindextext

%
\IEEEpeerreviewmaketitle

\section{Introduction}
Matrix factorization techniques can effectively detect defects in sequence of thermal images \cite{r2,r4,r5,r6,r12,r43,r18,r50,r10,r51,r52,r53,r54,r55,r56}. Some well-known matrix factorization methods, such as principal component analysis/thermography (PCA/PCT) \cite{r2} and non-negative matrix factorization (NMF) \cite{r1,r52,r56}, have been used for past several years in different fields. However, the major challenge of dealing with high-dimensional infrared imaging sequences in different applications persists. In this paper, we propose a comparison assessment of low-rank matrix approximation by applying semi-, convex-, and sparse-NMF to two different scenarios through active and passive thermographical imaging.\\
After using PCT to detect thermal defects \cite{r1},  several other alternative approaches have been used to modify the PCT to improve its performance by using fixed eigenvector analysis method \cite{r3x1}, incremental PCT \cite{r7}, or candid covariance-free incremental principal component thermography (CCIPCT) \cite{r4,r51}; these approaches resulted in better computational load by a fixed set of previously generated eigenvectors and covariance free approach. Sparse- PCT \cite{r4x,r5,r6} (or even presented as sparse dictionary matrix decomposition \cite{r16}) considers regularization terms to increase the sparsity in the analyses and strengthening the important bases to detect defective patterns.

\begin{figure*}[!t]
\centering
\includegraphics[width= 0.95 \textwidth, angle=-00]{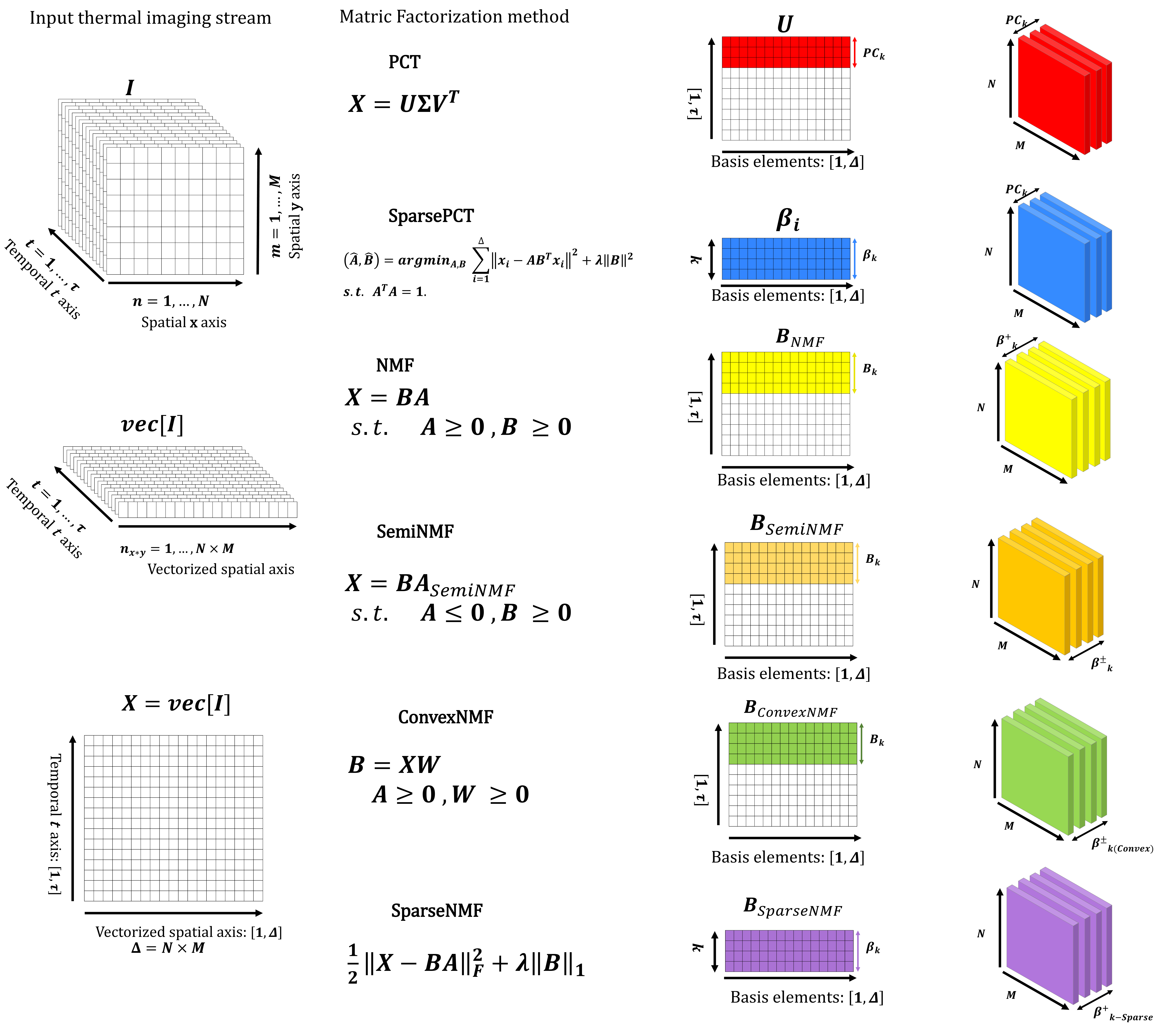}

\caption{Workflow of matrix factorization methods to obtain low-rank estimation of input thermal imaging stream.}
\label{figR1}
\end{figure*}

Nevertheless, all these methods successfully followed the fundamental principle of thermal non-destructive testing (NDT) signals and considered the initial eigenvector as the predominant bases of the entire data set. A spatial-transient-stage tensor mathematical model of the Tucker decomposition algorithm motivated by eigendecomposition is used in an inductive thermography system to track and characterize variations in properties \cite{r14}. This method has also been expanded to show spatial-sparse thermal patterns and involves electromagnetic energy along with thermography \cite{r15}; the method has considerable similarity to a non-negative pattern separation model by using sparse greedy-based principal component analysis (SGPCA) \cite{r16}. The reconstruction of input thermal sequence involves a combination of negative and positive bases due to lack of restriction in extracting bases and coefficients in PCT, which might cause overlapping among the distinct basis in low-rank matrix approximation.
Non-negative matrix factorization (NMF) is a matrix factorization technique \cite{r3} similar to PCA and has additive constraints in basis and coefficient matrices; this method decomposes an input matrix into low-rank non-negative basis \cite{r3} to solve the aforementioned issue. NMF is used to estimate geometrical properties along with PCT and archetypal analysis (AA) \cite{r12}. An approach for dimensionality reduction, which is not an IRNDT method, is performed using NMF for visible and infrared fusion \cite{r17}. NMF is applied in IRNDT through two ways of computations using gradient descend (GD) \cite{r20} and non-negative least square (NNLS) \cite{r21,r22} to evaluate cultural heritage objects and buildings; this method exhibits considerable performance for detecting subsurface defects \cite{r19}. An ensemble joint sparse low-rank matrix decomposition is presented for detecting thermal defects in CFRP specimens by using the optical pulse thermography (OPT) diagnosis system \cite{r38}. Semi-NMF and Sparse-NMF for thermography very briefly discussed an application of these approaches without detailed analyses \cite{r42,r43}, without discussing Convex-NMF.  Despite considerable developments in detecting subsurface spatio-thermal defect patterns, semi-, convex-, and sparse- NMF have not been discussed yet. Moreover, the effects of non-negative constraints have not been investigated in thermography applications
Changing the bases in the NMF method to convert the problem to a segmentation problem while selecting a predominant basis remains challenging. This study shows an application of low-rank semi-, convex-, and sparse-NMF in thermography and breast cancer screening to detect defects and identify symptomatic patients, respectively. 
The rest of the paper is organized as follows. In the next section, thermal transfer for imaging system in active and passive thermography is summarized. In Section III, the methodology of the approach will be briefly described by applying semi-, convex-, and sparse-NMF analysis for thermography. The experimental and computational results are presented in Section IV, and discussion is shown in Section V. The conclusions are summarized in Section VI.

\section{Thermal transfer for imaging system}
Thermal camera captures the spatial variation of temperature on the targeted region of interest (ROI) on specimen over the time. This heat transient can be through active or passive thermography procedure. In general, thermal transfer/heat conduction equation of a specimen can be summarized by following equation:

\begin{equation}
 \frac{\partial T}{\partial t} = \frac{k}{\rho C_\rho} \left( \frac{\partial^2 T}{\partial x^2} + \frac{\partial^2 T}{\partial y^2} + \frac{\partial^2 T}{\partial z^2} \right) + \dot{q}
\end{equation}\\
where $T = T(x,y,z,t)$ is a temperature field, $k$ is thermal conductivity constant from the material $(W/m .K)$. $\rho$ is the density ($kg/m^3$), $C_p$ is specific heat $(J/kg .K)$, $\dot{q}(x,y,z,t)$ is the internal heat generation function per unit volume, in the passive thermography ($\dot{q}\cong 0 $ in active thermography). Other thermal transfer (i.e. convection) through the object and the fluid were neglected here.  
Applying infrared thermography on biological organs and tissues composed of fat, blood vessels, parenchymal tissues, and nerves increase the complexity of the previous equations. This complexity aggravates with some degree of uncertainty regarding the rate of blood perfusion and metabolic activity. Pennes’ bioheat equation \cite{r46,r48} provides accurate thermal computations and states as follows:
\begin{equation}
 \rho C_\rho \frac{\partial T}{\partial t} = \bigtriangledown .(k \bigtriangledown T) + \omega_b C_b (T_a-T) + \dot{q}
\end{equation}\\
Where $\omega$ represents the flow rate of blood, $\dot{q}$ is the metabolic rate (heat generation), and $b$, and $a$ in $\omega_b C_b (T_a-T)$ additive term stand for blood, and arteries (in targeted tissue-breast), respectively.

\section{Method}
\subsection{Related works}
Let $I$ represents a sequence with $\tau$ thermal images where each image has spatial dimension of $N \times M$. The $X$ is the input data constructed by appending the vectorized thermal images, $X = \{x_1,x_2,\dots, x_{\tau}\}$ and $x_1 = vec[I_1]$. The PCT \cite{r1} decomposes the input data, $X$, to $U \Sigma V^T$, where $U$ and $\Sigma$ represent the eigenvector (basis) and eigenvalue (coefficient) matrices. The bases corresponding to 80$\%$ highest variance capture the maximum thermal patterns (maximum thermal heterogeneity) among the bases, called low-rank matrix approximation. However, there is no guarantee of independency and non-overlapping property in the decomposed bases, which increases collinearity among the bases. This leads to generate many similar components and difficult selecting the representative basis. This considers to be a drawback of PCT despite a substantial contrast respond of the algorithm in low-rank approximation.
Sparse-PCT \cite{r4x,r5,r6} optimizes the selection of the bases by adding penalty terms, that increases the sparsity in the bases (from the basis matrix $B$ - Figure 1) and moderates the collinearity by restricting the domain of solution. Thermal patterns measured by low-rank matrix approximation using Sparse-PCT show more separability among the bases compare to PCT, despite more computation load for its calculations \cite{r4x,r5,r6}. This subsequently lead to ambiguity among the bases while extracting the components.
Similar to PCT, the NMF algorithm can be presented by a linear combination of $k$ basis vectors to reconstruct the data, $X$, whereas limited the overlapping bases by having additive non-negative constraints for coefficient and basis matrices. NMF is comparable to the PCA decomposition yet the PCA’s basis vectors can be negative. 
Input data $X$ can be shown by a linear combination of $\tau$ bases, $B = [\beta_1,\beta_2,\dots,\beta_{\tau}]$ and $A$ coefficient. This is showed as follows:

\begin{equation}
    \centering
    X = B A  \:\:\:\:\:\: s.t.  \:\:\:\:\:\: A \geq 0 \:\:\:\:\:\:   B\geq 0 
    \label{eq.2}
\end{equation}
where $X \in R_{\Delta \times \tau}^+$, $\Delta = MN$, $A \in R_{\tau \times \tau}^+$, and $B \in R_{\Delta \times \tau}^+$ or with $\ell_2$ equation that provides Euclidean distance \cite{r3} (Figure 1):

\begin{equation}
    \centering
    min_{A,B} f(A,B) = \|X - BA \|^2 \:\:\:\:\:\: s.t.  \:\:\:\:\:\: A \geq 0 \:\:\:\:\:\:   B\geq 0 
    \label{eq.3}
\end{equation}

NMF assumes that matrices $X$, $B$, and $A$ are not negative. This property significantly increases the chance for having bases with unique characteristic of thermal patterns as each basis corresponds to specific thermal patterns captured through selecting the direction of the highest thermal variance. Hence, the NMF directly associates with clustering due to non-negative constraints of bases \cite{r3}. NMF used in thermography and $k$ first bases could represent thermal patterns using gradient descent and non-negative least square error algorithms. Restricting both coefficients and bases in the NMF might limit the selection of the bases, and lead to discarding useful information. This study challenges this problem by using semi- and convex- NMF, where the constraints are loosen to determine their effects on selected bases.

\subsection{Semi-, Convex- and Sparse- NMF in thermography}
Semi-NMF performs matrix decomposition while $A$ is restricted to be non-negative but there is no restriction for matrix $B$. This provides freedom to the basis matrix (similar to PCT) which is restricted by coefficient matrix (like NMF). Having constrained coefficients control the bases avoiding over-expression of collinear components. Semi-NMF has no constraint for basis vectors, which provides limited freedom to the selected bases to capture all the possible heterogeneous thermal patterns. Similarly, convex-NMF imposes the constraint of basis vectors lie within the column space of $X$:
\begin{equation}
   \beta_\ell= w_{1\ell} x_1+ w_{n\ell} x_2+\dots+w_{MN\ell} x_MN = X w_{\ell}, \:\:\:\:\:\:  B = XW
\end{equation}
Where there is a convex combination of the columns of $X$. This restriction has the benefit that we could take the columns $\beta_l$ as weighted summations of particular data points. In fact, these columns would motivate a concept of centroids and this restricted form of $B$ factor refers as Convex-NMF \cite{r31}.
Low-rank matrix approximation using semi- and convex- NMF provide more freedom to bases as compare to NMF while controlled by non-negative coefficients. It means despite potentially negative bases, they do not overlap. This property makes these algorithms to perform grouping of bases driven by their thermal variability (similar to K-means clustering \cite{r31}) on input thermal data, $X$.
The aforementioned minimization problem (3) does not have concurrent convex property for both matrices, $A$, $B$, whereas the problem is convex for each matrix discretely. Many researches have proposed ways and optimization solution such as: GD algorithm and NNLS by Paatero and Tapper \cite{r9}, multiplicative algorithm by Lee and Seung \cite{r3}, a computational improved GD algorithm by Lin \cite{r20}, and modified alternating non-negative least squares (ANNLS) by Liu et al. \cite{r21}. To compensate the uniqueness of the decomposition and enforcing a representation of basis, sparseness constraints are proposed for NMF \cite{r22}. A $\ell_1$ norm penalty term has been introduced by implementing the following equation:

\begin{equation}
    c_{Sparse-NMF} = 1/2 \|X-AB \|^2_F + \lambda \|B\|_1 \\
 \:\:\: 
    \label{eq.4}
\end{equation}\\
Let $\|B\|_p$ is the $\ell_p$-norm of $B$ given by $L_p = (\sum_{d,m} \|B_{d,m} \|^p )^{1/p}$ similar to usual $\ell_1$ penalty term to imitate the $\ell_0$ behavior \cite{r27} to calculate $B$ for convex $A$. It is the unconstrained least squares minimization with $\ell_1$-norm constraint, also referred as least absolute shrinkage and selection operator (LASSO) \cite{r28}. The solution for all values of $\lambda$ achieves through applications of the least angle regression and selection algorithm (LARS) \cite{r30}. 
Similar to sparse-PCT but with non-negative constraints, low rank sparse-NMF obtains by selecting $k$ bases correspond to the highest coefficients (as showed in Figure 1). Sparse constraint in the calculation alleviates noisy thermal signal in the bases, which provides more robust thermal patterns in the region of interest (ROI).
Sparse-NMF theoretically more suitable for extraction of thermal patterns as it does have minimal collinearity among bases, which alleviates the ambiguity in selecting targeted components. Moreover, sparsity in the calculation of bases, which decreases noise effect in measurement of low-rank matrix approximation.

\section{Results}
The proposed procedures for thermal pattern detection were examined. The methods were applied for analysis of three different surface panels with diverse material and thermal breast cancer screening datasets. The resulting semi-, convex- and sparse- NMF were then compared with other state-of-the-art data post-processing algorithms. The results were assessed in the presence of additive Gaussian noise to determine the stability of methods facing noise by calculating signal-to-noise ratio (SNR).

\begin{table}[!t]
	\begin{center}
	\caption{Clinical information and demographics of the database for mastology research.}
			\includegraphics[width=0.52 \textwidth, angle=-00]{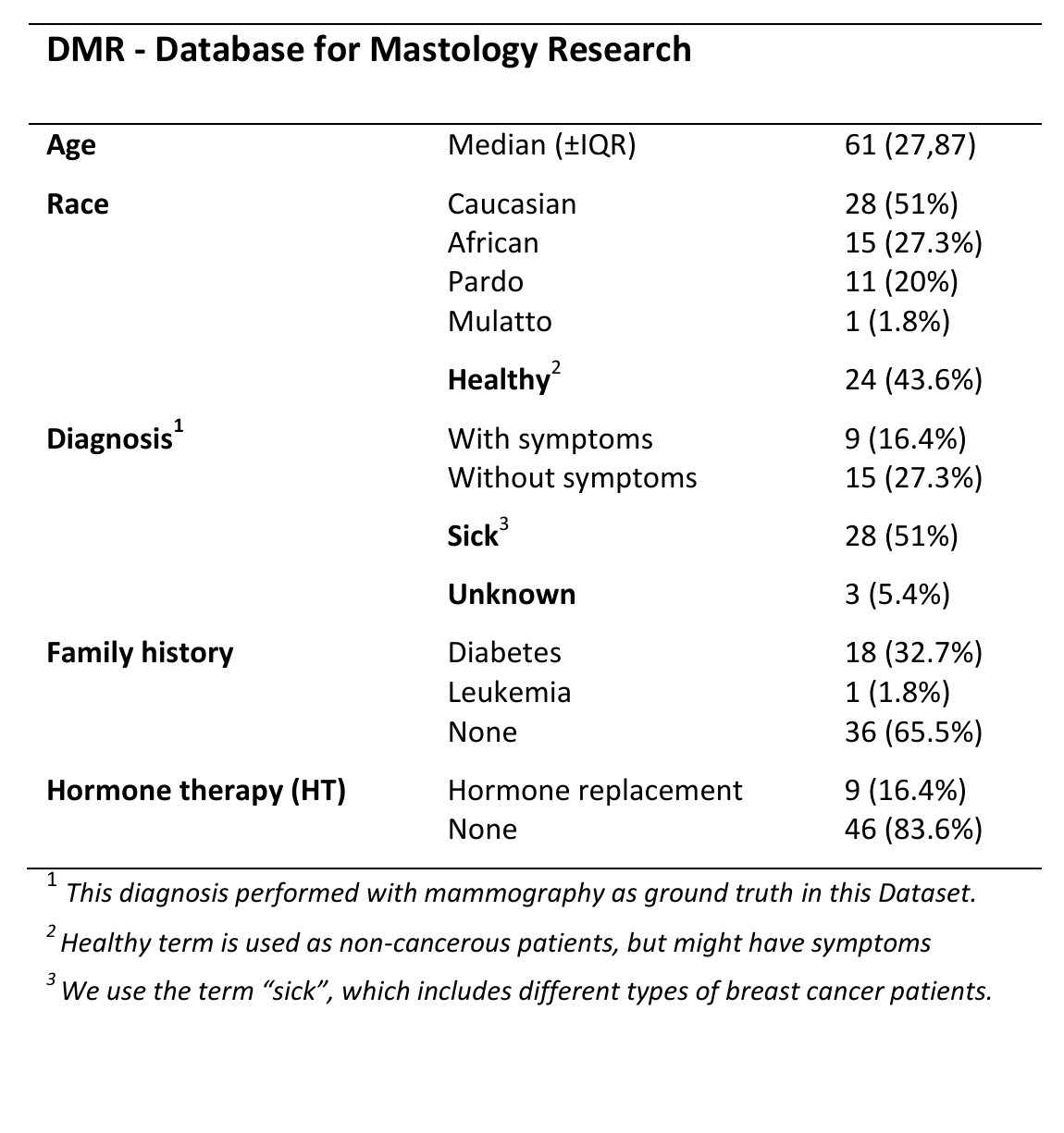}
		\label{tabl1}
	\end{center}
\end{table}

\subsection{Infrared image sets}
\subsubsection{IR-NDT}
The inspection results were obtained from the front side of the specimens, namely, carbon fiber-reinforced polymer composites (CFRP), Plexiglas (PLEXI), and aluminum (AL) plates. The inspected plates had some defects in different depth ranges: 0.2–1 $mm$ for CFRP plate, 1–3.5 $mm$ for Plexiglas, and 3.5–4.5 $mm$ for aluminum. Two photographic flashes were used: 5 $ms$ Balcar FX 60 and 6.4 kJ/flash thermal pulse. The infrared camera used was Santa Barbara Focal plane (MWIR, nitrogen cooled, InSb, $320 \times 256$ pixels). The acquisition parameters were set as follows: sampling rate, $f_s$ =157 $Hz$; Duration, $t_acq$ = 6.37$s$; time step, $D_t$ = 0.025$s$; truncation window, $w(t)_s$= 6.37 $s$; and total number of frames of 250. The sampling rate was 157 $Hz$, and 1000 images were recorded. In two other image sets, the experimental setup was similar to the previous experiment of inspecting photographic flashes and infrared camera \cite{Ref16}. 
AL and PLEXI specimens were tested for defect detection and estimation of the depth of each defect. The Plexiglas plate has a thickness of 4$mm$ and contains six artificial defects (flat-bottom holes) located at depths of 3.0, 4.0, 3.5, and 4.5 $mm$ (Figure 2).

\begin{figure}[!t]
\centering
\includegraphics[width= 0.5 \textwidth, angle=-00]{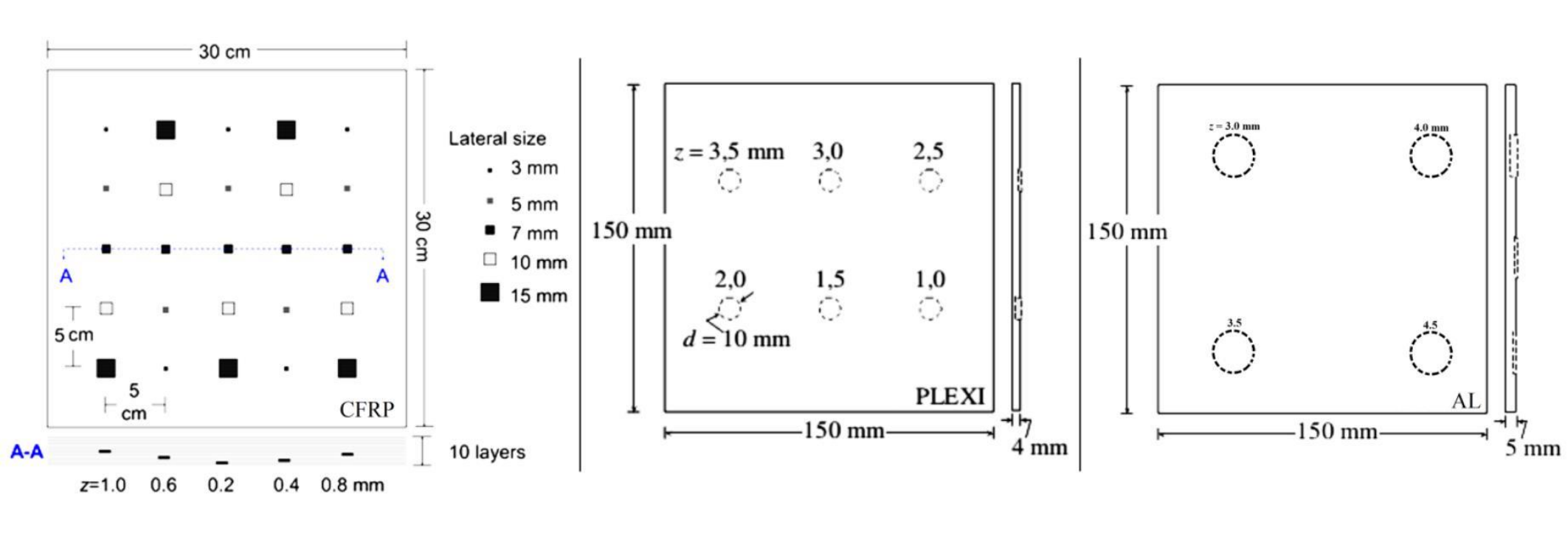}

\caption{\textbf{Schematic of the specimens used for benchmarking of the proposed approach.} The picture in the first column represents carbon fiber-reinforced plastic (CFRP) specimen with 25 defects of different depths and sizes. The sizes and depth of each defect is clearly indicated. The second and third columns present PLEXI and AL specimens with six and four defects, respectively, where each defect is of a different size and depth.}
\label{figR2}
\end{figure}

\subsubsection{Infrared breast cancer dataset}
Fifty-five participants who were healthy (with/without symptoms) or sick (diagnosed by mammography) were employed for breast screening. The median age in our study sample was 61 years, and the participants comprised 28 Caucasian (51$\%$), 15 African (27.3$\%$), 11 Pardo (20$\%$), and 1 Mulatto (1.8$\%$) women. Among the participants, 18 had history of diabetes in their families (32.7$\%$), and 9 were undergoing hormone replacement (16.4$\%$). All patients had IR images obtained by the following acquisition protocol: images have spatial resolution of 640 $\times$ 480 pixels and were captured by a FLIR thermal camera (model SC620) with sensitivity of less than 0.04$^\circ C$ range and capture standard of −40$^\circ C$ to 500$^\circ C$ \cite{r39}. Table 1 shows the clinical information and demography of the cohort.

\begin{figure*}[!t]
\centering
\includegraphics[width= 0.5 \textwidth, angle=90]{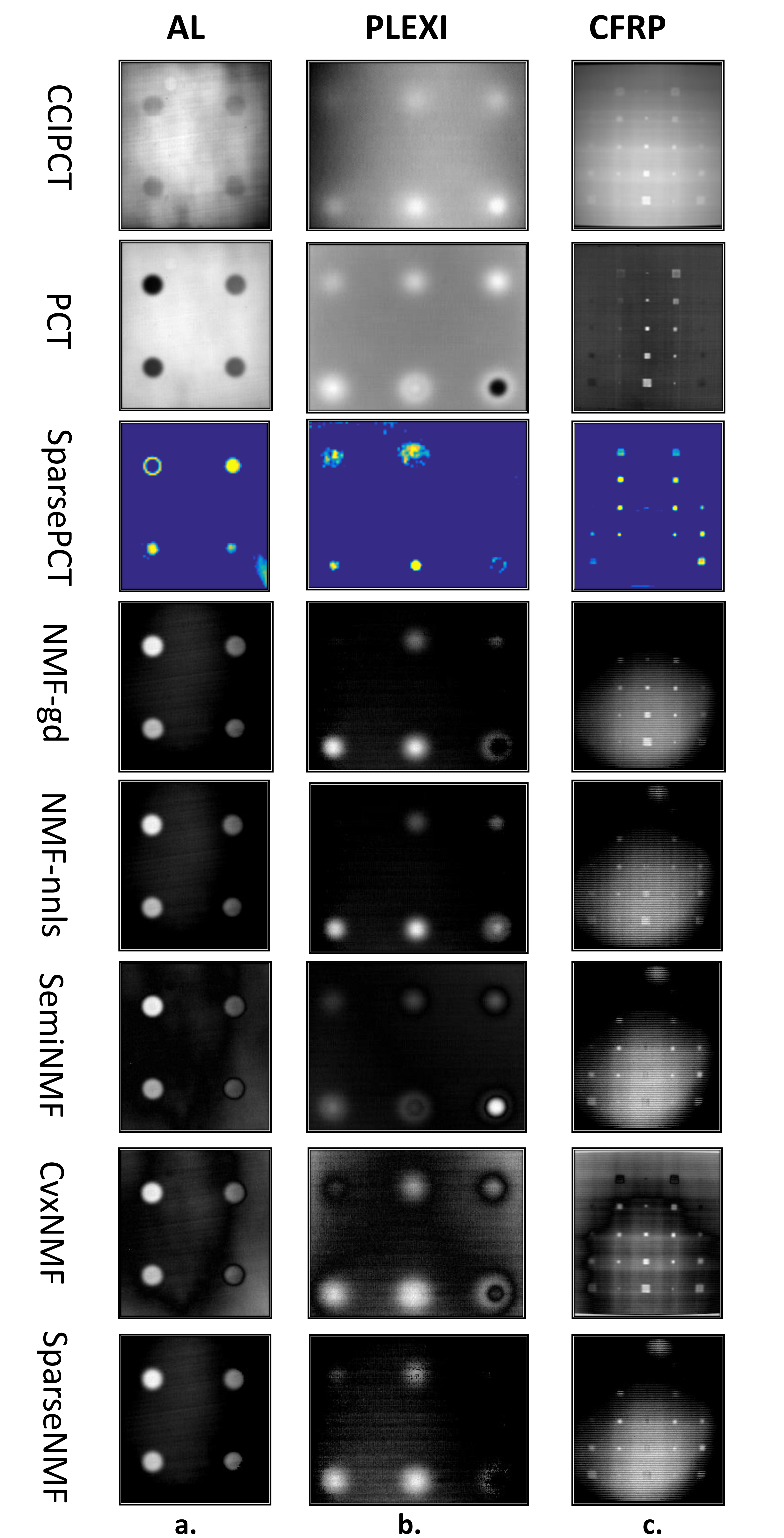}

\caption{\textbf{(a-c)} Low-rank matrix factorization for aluminum, Plexiglas, and CFRP plates for active thermography. }
\label{figR3}
\end{figure*}

		

\begin{table}[!b]
\caption{Jaccard accuracy index of each algorithm for all specimens in different depths and sizes of defects.}
\includegraphics[width=0.5\textwidth,angle=-00]{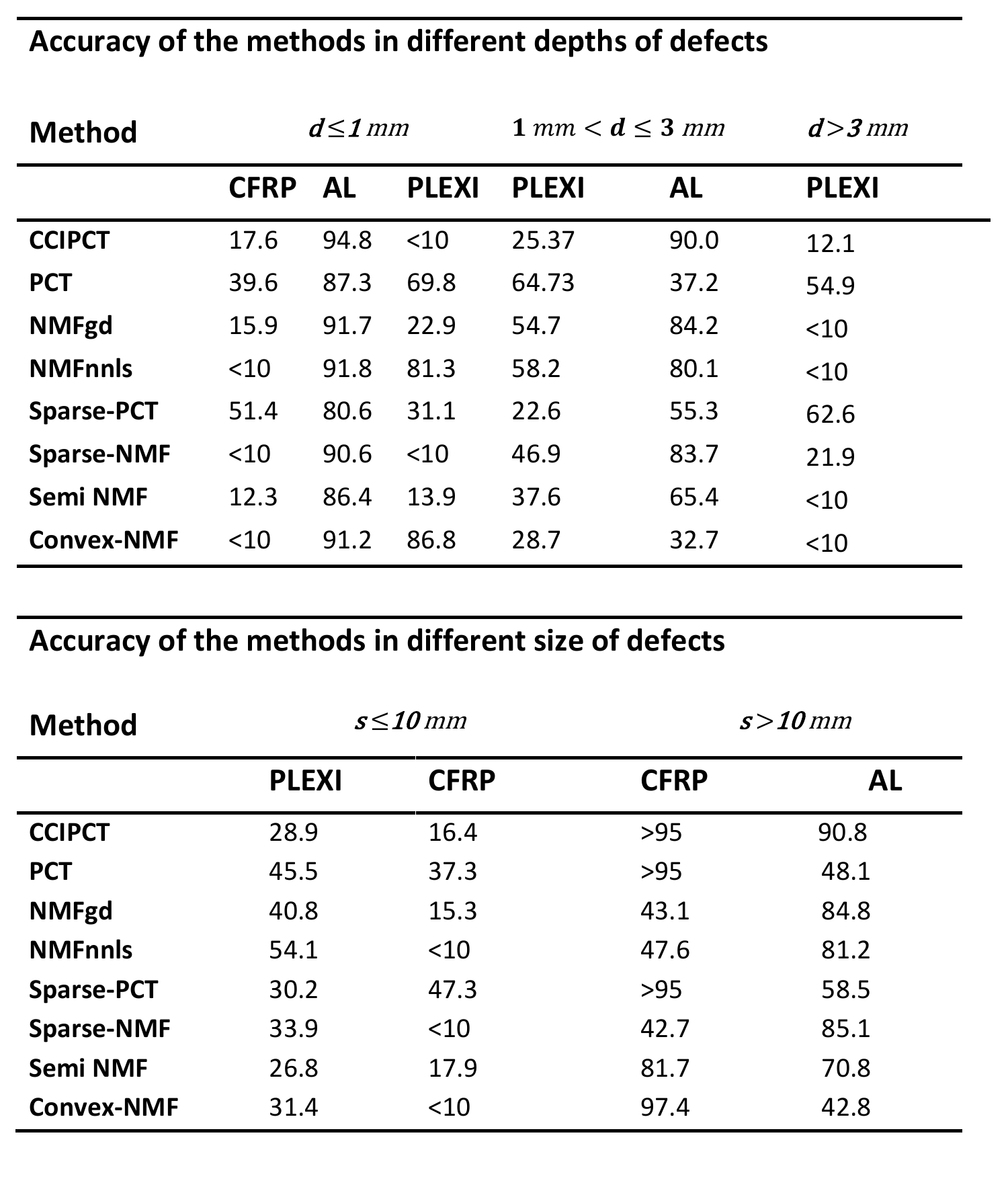}
\label{table2}
\end{table}

\subsection{Outcome of thermal defect patterns in IRNDT}
To calculate the quantitative accuracy, we employed a binary image as ground truth (GT) and as a reference. The GTs were labeled the pixels in defects and background of the specimens by 1 and 0 intensity levels, respectively. A metric called Jaccard index was used to quantify the percent overlap between the GT and our predicted output and was measured by the number of pixels shared between the GT and resulting masks divided by the overall number of pixels across both masks (or area of union). Given that the binary results were compared to the binary reference, a threshold would be used to convert the results to binary images. In this regard, we empirically selected the optimum threshold based on the correctness of subsurface detection (as the location of defects are known).
\begin{equation}
    J(W,GT) = \frac{|W \cap GT|}{|W \cup GT|}
\end{equation}
where $W$ and $GT$ are the binarized low-rank matrix approximation of thermal sequence and GT, respectively. Figure 3 presents the results of the subsurface defect detection using semi-, convex-, and sparse-NMF compared with state-of-the-art approaches, such as PCT \cite{r2}, candid covariance-free incremental principal component thermography (CCIPCT) \cite{r4}, NMF-gd, NMF-nnls \cite{r19}, and sparse-PCT \cite{r5}. The qualitative results indicated the accuracy of semi-, convex-, and sparse-NMF relative to the state-of-the-art factorization methods. The results of semi-, convex-, and sparse- NMF relatively indicate more discriminative patterns between defect and background for all three specimens, where sparse- NMF, similar to Sparse- PCT, showed lesser noisy approximation.
\begin{figure*}[!htb]
\centering
\includegraphics[width= 0.95 \textwidth, angle=-00]{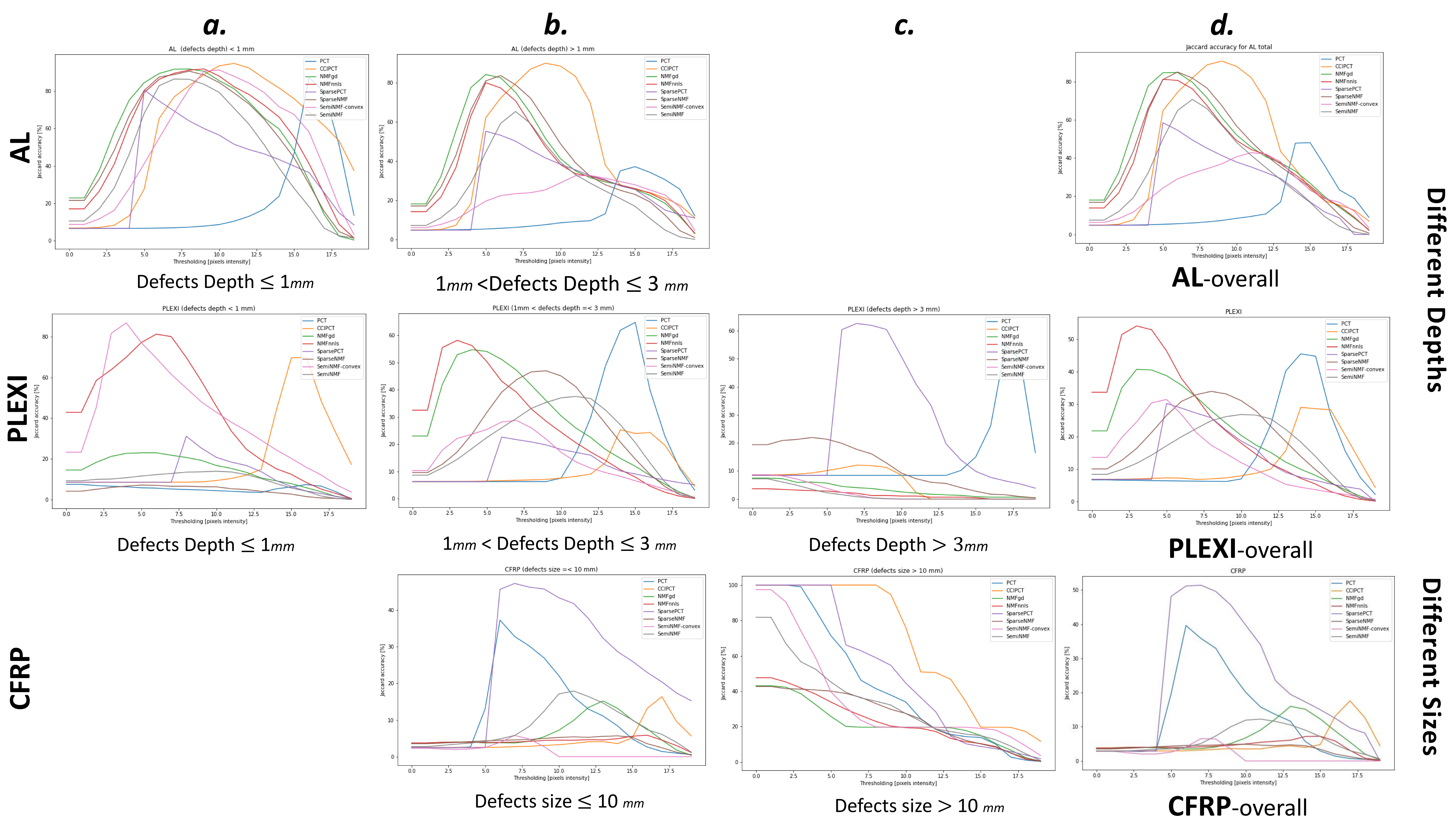}

\caption{Different depths and sizes of the defects for three specimens. To estimate such accuracy, we changed the pixel intensity by 5$\%$ in each step (X-coordinate), applied different binarized thresholds, and calculated Jaccard index accuracy. Columns a-c represent the accuracy of defect detection for depths and sizes with the respect to the type of specimens.}
\label{figR4}
\end{figure*}
\subsection{Depth and size of the defects}
A thresholding should be applied to the approximated matrix to calculate a binary low-rank matrix approximation using the presented approaches. Figure 4 and Table 2 show the effect of different defect sizes and depths on the overall accuracy of detecting different defects. Moreover, Figure 4 represents changes in the threshold by using different pixel intensities (5$\%$ variation in each step) and its effect on the Jaccard accuracy index. Table 2 indicates the highest accuracy for each case by changing the pixel intensity. 

In the case of aluminum, Tabel 2 presents the quantitative assessment of semi-, convex-, and sparse-NMF in comparison with the state-of-the-art thermographic approaches and showed 70.8$\%$, 42.8$\%$, and 85.1$\%$ accuracy with 27.5$s$, 1$s$, and 39.52$s$ computational load, respectively. The CCIPCT and PCT had Jaccard index range of 90.8$\%$ ($<1s$) and 48.1$\%$ (0.25$s$), respectively. NMF-gd and NMF-nnls were accurate, with values of 84.8$\%$ (14.81$s$) and 81.2$\%$ (45.18$s$), respectively. Overall, big defects and middle-depth defects ($1 mm<d \leq 3 mm$) showed better accuracy than the other defects (Table 2). 

For the CFRP specimen, the overall range of the Jaccard accuracy was lower due to bigger defects, thereby affecting the detection of spatio-thermal defect patterns (Figure 3.c). PCT, CCIPCT, and sparse- PCT exhibited the highest accuracy, that is, $>95\%$ (1$s$, 7.2$s$, and 63.7$s$), whereas convex-NMF and semi-NMF had 81.7$\%$ and 97.4$\%$ accuracy within 17.53$s$ and 286.57$s$, respectively (for $s>10 mm$). Sparse-NMF showed low accuracy due to hard tuning of the regularization parameters to obtain basis elements in the decomposition; as such, the thermal defective pattern were poorly detected (< 10$\%$) (Table 2).
PLEXI showed higher accuracy for less deep defects than the other depths, and Convex-NMF and NMF-nnls showed the two highest accuracies (86.8$\%$- 9.37$s$ and 81.3$\%$- 96.33$s$, respectively). Overall, the average of defect detection for PLEXI reached 36.45$\%$ through Jaccard index measurement (Table 2 and Figure 8).

\subsection{Results of thermal breast cancer screening}
Seven low-rank matrices were extracted from every participant, where each case had 23 thermal sequences. Out of 7 low-rank matrices, one matrix manually selected based on having relatively the best contrast property.
The selected representative images are shown in Figure 5 for each participant. Heterogeneous thermal patterns were detected for 31 participants in breast cancer screening indicating sickness or abnormality (healthy participants but with symptoms) in the breast area (see Figure 5.a-c). More homogeneous thermal patterns were detected among the healthy participants (healthy with no symptoms, Table 1, Figure 5.d-f). 

\begin{figure*}[!t]
\centering
\includegraphics[width= 0.85 \textwidth, angle=-00]{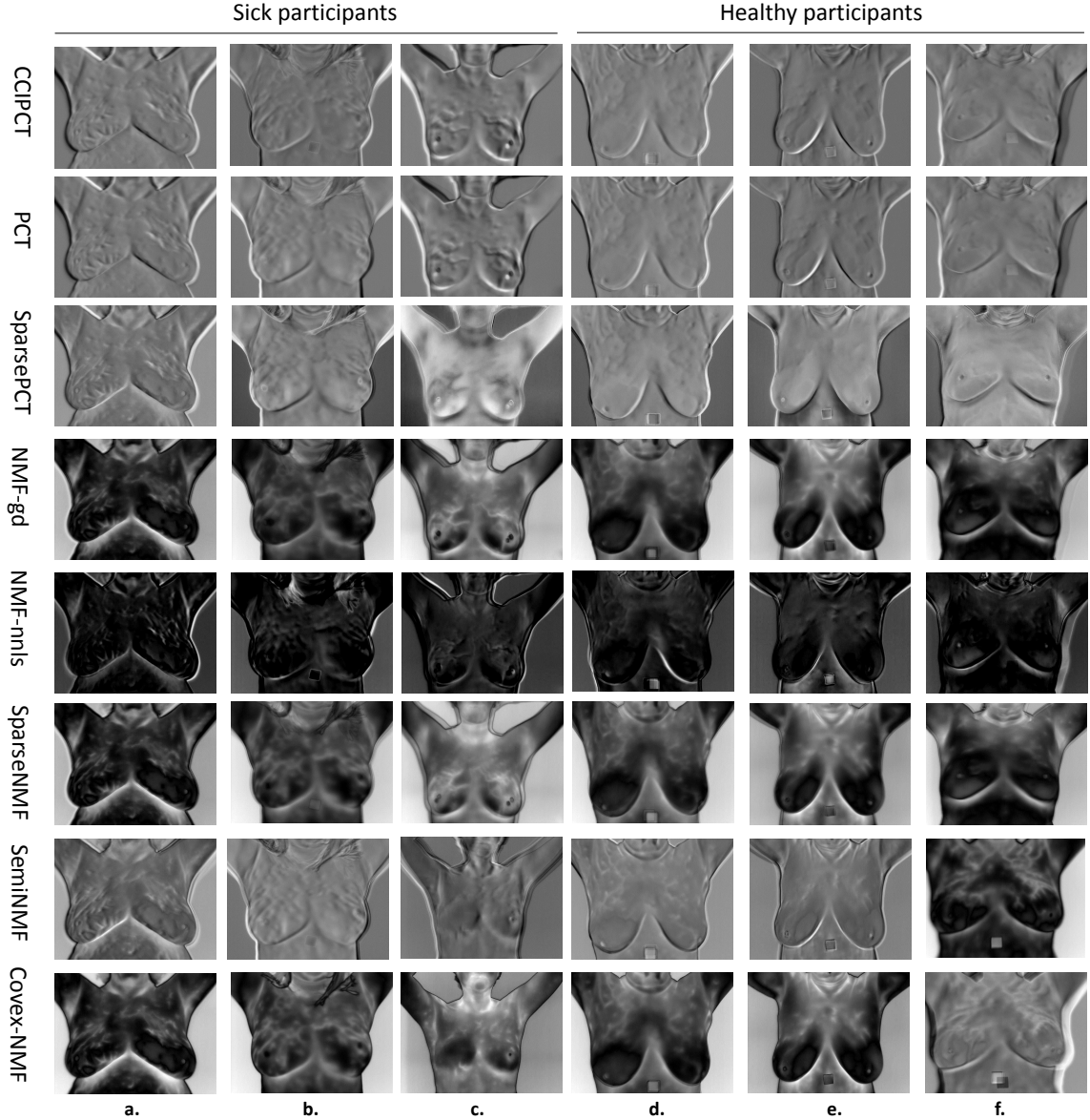}

\caption{Low-rank approximation of thermal sequence determined using different matrix factorization techniques. Columns (a-c) show symptomatic patients (diagnosed by mammography as cancer patients or healthy with symptoms), whereas columns (d-f) show the result of methods for healthy participants.}
\label{figR5}
\end{figure*}

\subsection{Outcomes of thermal breast cancer screening}
Statistical analyses were performed to measure the thermal heterogeneity of the breast area by using texture analyses. The thermal measurements were encoded in the gray level intensity. The thermal level co-occurrence matrices (TLCMs) of the breast area were calculated for the ROI \cite{r40}. For each patch, TLCM with a horizontal offset of 4 (two distances $(0,1)$ and two angles $[0,\pi/2]$) were computed to capture thermal patterns in the ROI. 

\begin{figure*}[!t]
\centering
\includegraphics[width= 0.85 \textwidth, angle=-00]{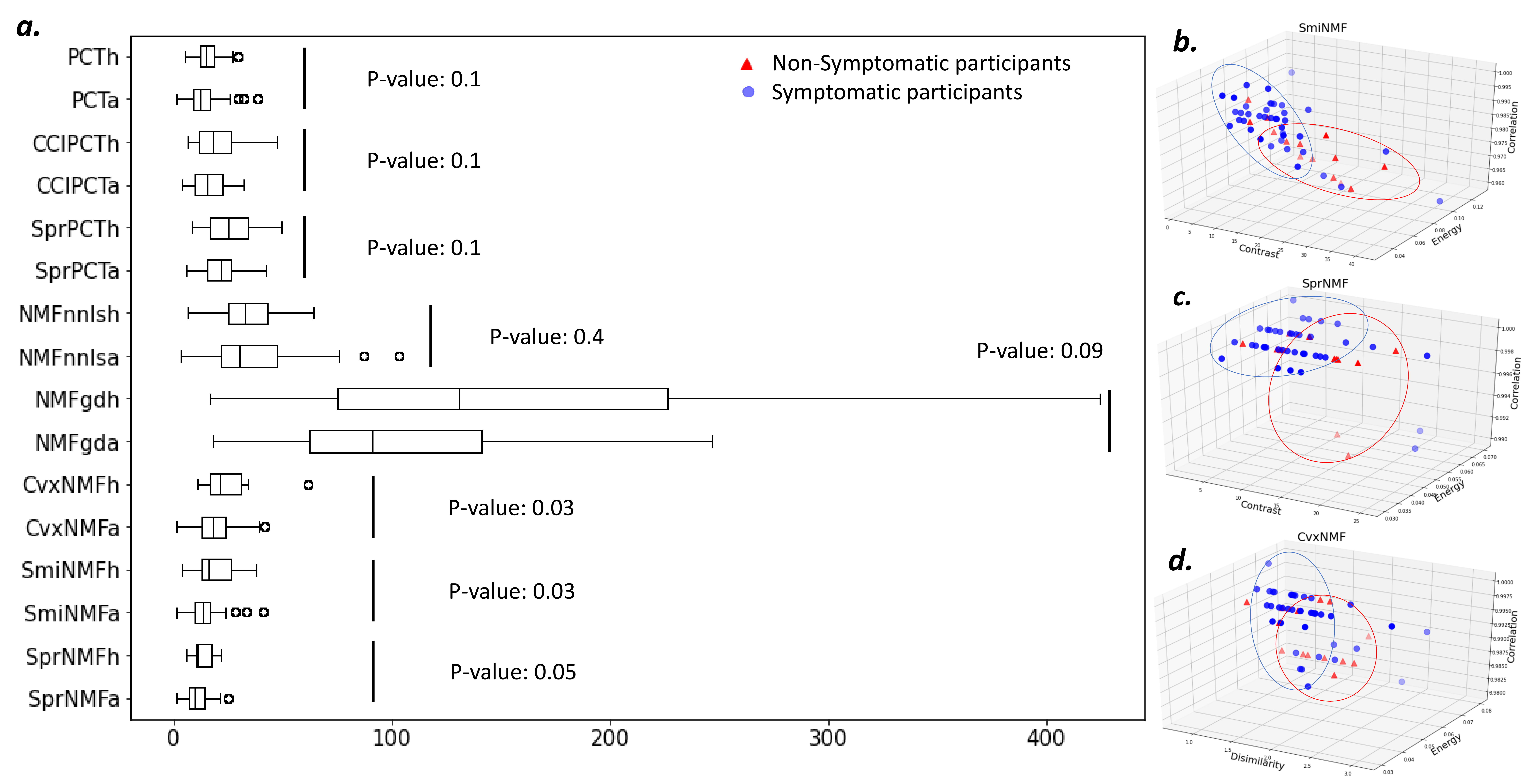}

\caption{Stratification of participants into symptomatic (abnormal-“a” suffix) and non-symptomatic (healthy-h suffix) groups for each matrix factorization algorithm as presented by boxplots and Kurskal Wallis test (a.). b-c represent the 3D views of grouping using TCLM’s properties.}
\label{figR6}
\end{figure*}

\begin{figure}[!t]
\centering
\includegraphics[width= 0.5 \textwidth, angle=-00]{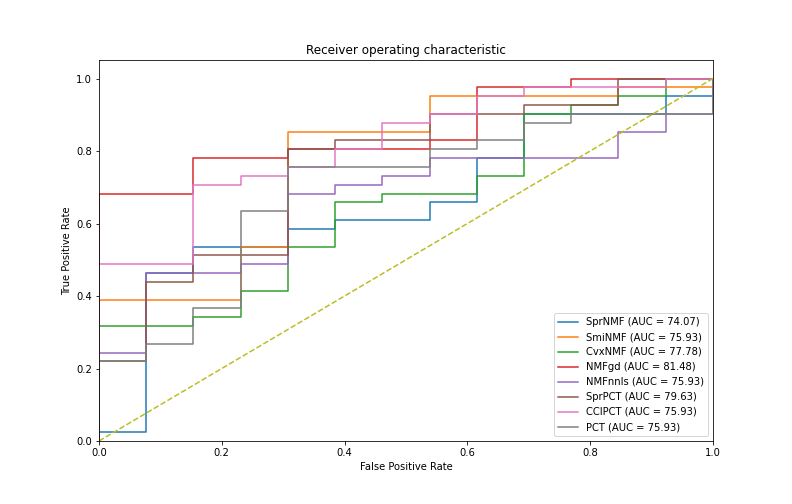}

\caption{ROC graph for different matrix factorization approaches for multivariate covariate logistic regression binary classification (abnormal or healthy participants).}
\label{figR7}
\end{figure}
Four properties of the TLCM matrices were determined to measure the level of contrast, dissimilarity, correlation, energy, and homogeneity among the pixels in the ROI by using the following equations:\\
Contrast: 
\begin{equation}
     \sum^{levels-1}_{i,j=0} P_{i,j}(i-j)^2
\end{equation}
Dissimilarity: 
\begin{equation}
     \sum^{levels-1}_{i,j=0} P_{i,j}|i-j|^2
\end{equation}
Homogeneity: 
\begin{equation}
     \sum^{levels-1}_{i,j=0} \frac{P_{i,j}}{1+(i-j)^2}
\end{equation}
Energy:
\begin{equation}
     \sqrt{ \sum^{levels-1}_{i,j=0} P_{i,j}^2 }
\end{equation}
 Correlation:
\begin{equation}
    \sum^{levels-1}_{i,j=0} P_{i,j} \left[  \frac{(i-\mu_i)(j-\mu_j)}{ \sqrt{(\sigma_i^2)(\sigma_j^2)}}\right]
\end{equation}
We stratified the participants based on these properties and compared with the binary GT data obtained from mammography information (symptomatic or non-symptomatic). Kurskal Wallis test was performed to determine the statistical difference between two groups (symptomatic versus non-symptomatic) of participants. 
Semi-, Convex-, and sparse- NMF showed statistically significant separation of the two groups of participants when split on contrast-based TLCM ($p = 0.03$, Figure 6.a). Sparse- PCT, PCT, CCIPCT, and NMF-nnls did not show strong stratification ability because they were not significantly discriminated ($p > 0.1$). NMF-gd showed slight separation strength, which was not significant ($p = 0.09$, Figure 6.a). The plot of the TLCM properties for the participants also showed potential separability between the groups (Figure 6.b-d), which confirmed the results of boxplots for separating participants (Figure 6.a). 
A logistic regression model fitted for multivariate thermal covariates (contrast, dissimilarity, correlation, homogeneity, and energy) was used to examine the hypothesis that the thermal heterogeneity can be used as a biomarker to stratify among participants (to determine symptomaticity). The accuracy levels were 74.1$\%$, 75.9$\%$, and 77.8$\%$ for Semi-, Convex-, and Sparse- NMF, respectively. The two highest accuracies were found to belong to NMF-gd and sparse- PCT, with values of 81.5$\%$ and 79.6$\%$, respectively. CCIPCT and PCT were commonly showed 75.9$\%$ accuracy (Figure 7). 

\begin{figure*}[!t]
\centering
\includegraphics[width= 0.9 \textwidth, angle=-00]{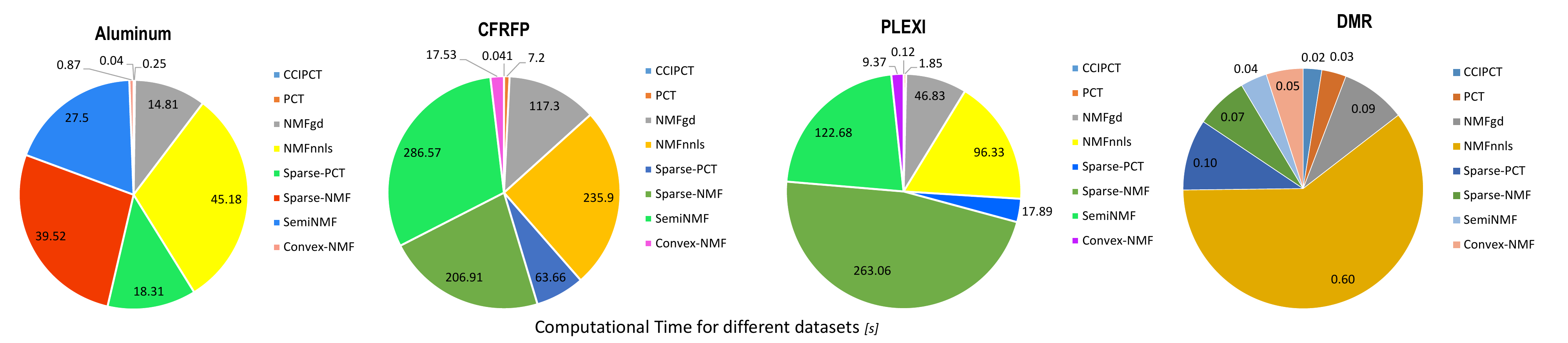}

\caption{Computational time for semi-, convex-, and sparse- NMF with the respect to other common methods in thermography.}
\label{figR8}
\end{figure*}

\begin{figure*}[!t]
\centering
\includegraphics[width= 0.9 \textwidth, angle=-00]{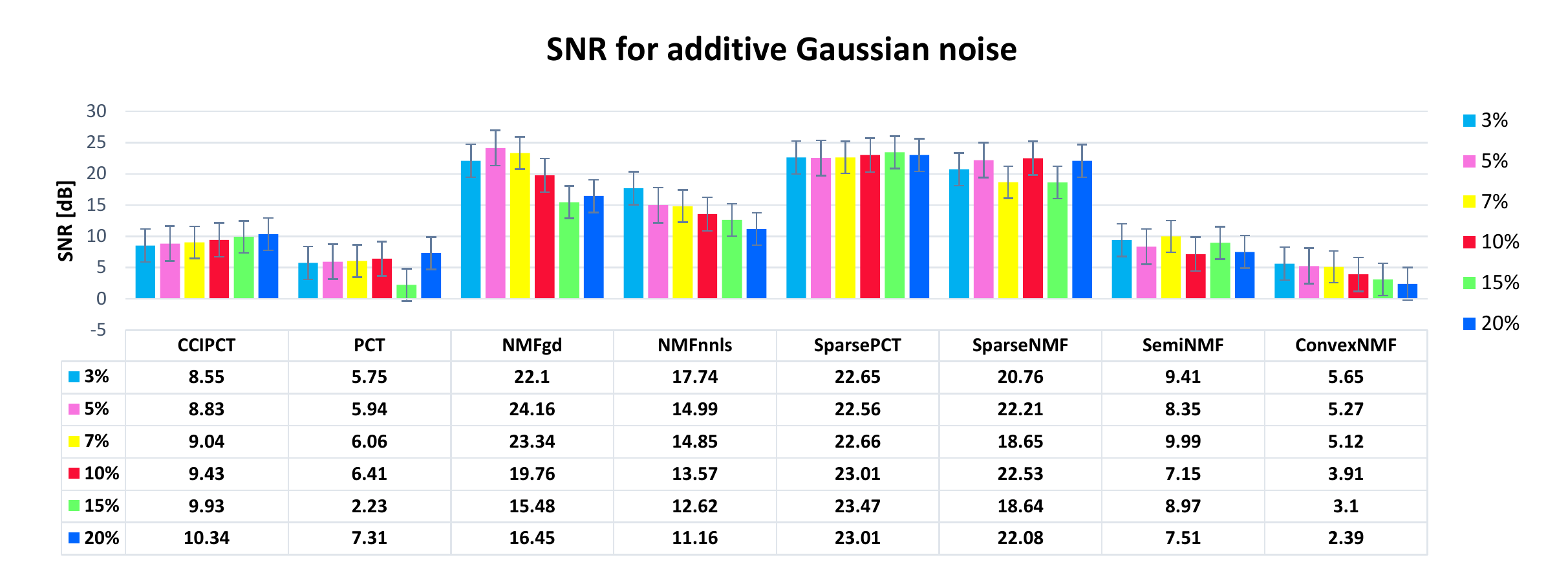}

\caption{SNR analysis of the state-of-the-art approaches in thermography.}
\label{figR8}
\end{figure*}

\subsection{Robustness in thermal defect patterns}
Noise in the sequence of thermal images is inevitable due to the nature of thermal acquisition. The sensitivity of acquisitions also depends on the acquisition’s conditions, which might intensify or mitigate the noise level in different systems. Here, the robustness of matrix factorization methods was evaluated while noise was added to the input signal. The results of semi-, convex-, and sparse- NMF were estimated by Jaccard index measure for additive Gaussian noise input. Table 2 represents the accuracy for semi-, convex-, and sparse- NMF compared with the state-of-the-art approaches. SNR for each method was calculated by the following equation (from \cite{r33}):

\begin{equation}
    SNR_3 = 10 \: \: log_{10} \frac{|\mu_S-\mu_N|^2}{\sigma_N^2}
\end{equation}
where $\mu_S$ and $\mu_N$ are the average levels of signal in the ROI and noise, respectively. $\sigma_N$ is the standard deviation of noise in the reference or sound region of the image \cite{r33}. Many other definitions have been established for the SNR. In this study, we follow one definition for the entire comparative analysis (following \cite{r33}).
In general, low-rank matrix factorization prevents additive noise in the spatio-thermal defect pattern detection due to low-rank noise reduction procedure in such algorithms. 
Here, the robustness results of each method were determined by the SNR metric, and Gaussian noise was added to the input image set. Low-rank matrix factorization analyses were performed to each case study while the noise level was increased from 3 $\%$ to 20 $\%$. 
The most robust methods in the presence of noise were sparse-PCT and sparse-NMF because of the additive regularization parameters and relatively nonlinear (and sparse) behavior of these methods. This finding might compensate for the low accuracy in the non-noisy environments and presents an advantage over the other state-of-the-art approaches. CCIPCT and Semi-NMF methods were more stable than SNR of input with higher additive noise; however, these methods had comparatively lower overall SNR than sparse-based methods. 
The computational process was partially performed with a PC (Intel Core 2Quad CPU, Q6600, 2.40 GHz, RAM 8.00 GB, 64-bit Operating System) by using MATLAB and Python programming language \cite{r41}.

\section{Discussion}
In this study, we used semi-, convex-, and sparse- NMF to extract thermal patterns for infrared diagnostic systems for active and passive thermography imaging. This study was designed based on the general trend of dimensionality reduction and defect detection methods (such as \cite{r2,r3,r4,r4x,r5,r6,r42}) and involved in-depth comparative analyses of specimens (for IRNDT) to identify potential patients with breast cancer (for DMR). 
In IRNDT, Convex- NMF showed higher accuracy than semi- NMF and Sparse- NMF for detecting defects (Table 2) possibly because the convex-NMF solutions are sparse and significantly more orthogonal unlike semi-NMF. 
In general both semi- and convex- NMF can give identical results similar to K-means clustering, where sharper clustering indicators can be given to convex-NMF. The restriction on basis can have significant effect on subspace factorization (creating more loss Equ.(5)), and more constrained results more degeneration of the accuracy \cite{r31}.\\
AL showed higher defect visibility in less deep defects in contrast to PLEXI. This finding might be due to the different thermal properties of the specimens with different depths. The overall performance of defect detection in the CFRP specimen was lower than that in aluminum and PLEXI due to its smaller defects ($1 mm>$, Figure 2). CFRP showed lower overall defect visibility due to less deep and bigger defects with the respect to different thermal properties. However, the results indicated that bigger defects in CFRP led to higher defect distinguishability. \\
The application of semi-, convex-, and sparse- NMF in low-rank matrix approximation is new in the field of thermography, but limited data are available for these approaches (\cite{r2,r3,r4,r4x,r5,r6,r42}). This study is the first to conduct such analyses for active and passive thermography. Moreover, comparative analyzing depths and sizes of defects for all the current state-of-the-art approaches increases the contribution of this study. \\
In DMR, Semi-, convex-, Sparse- NMF showed significant improvements in stratifying symptomatic patients from healthy participants (Figures 7, and 6.a) unlike other approaches (Figure 6.a). Moreover, convex- NMF showed higher accuracy than the results of IRNDT in finding heterogeneous thermal patterns, which might be due to the nature of passive thermography or better grouping thermal patterns because of better clustering property of convex-NMF than in IRNDT.
NMF-gd showed the maximum accuracy with slightly insignificant separability (Figures 6.a and 5), indicating the better performance of gradient descent algorithm over non-negative least square algorithm to optimize the $min_{B,A}f(B,A)$ . Thermal and infrared imagery has been used to determine the breast abnormality for past several years \cite{r44,r45}. Discussions about the better positions for such imaging acquisitions \cite{r46} and about the reliability of this modality for detecting the abnormalities \cite{r47} have been reported. However, the association of low-rank approximation of thermal heterogeneity with breast abnormality has not been discussed in literature and could have novel contribution to the field.\\
One limitation for applying the presented models is built based on manually selection of the bases with better defect visibility. Moreover, the low-rank approximation does not guarantee the best contrast in the predominant basis. In other word, the first rank approximation is not necessarily better that the second rank to show defects. Determining the internal parameters of each method (such regularization and number of iteration) was not discussed here and was selected based on the optimal performance of the algorithms because of comparative analyses among different methods and their validation.
The presented techniques offer some advantages. First, the problem can be more similar to clustering (particularly for active thermography) \cite{r31,r3,r8} due to the NMF's properties and selecting non-negative low-rank approximation, which could serve as unequivocal advantage on combining dimensionality reduction and detecting defects. Second, these methods considerably alleviate the effect of motion artifacts and noise, which can be substantial improvements in infrared thermography applications in medicine. To our best of knowledge, this study is the first to employ semi-, convex-, sparse- NMF methods for thermography in this extent.

\section{Conclusions}
This study proposed comparative analyses and new applications of semi-, convex-, and sparse- NMF algorithms in infrared diagnostic systems. We applied these approaches to detect defects in NDT specimens (\textit{i.e.} carbon fiber-reinforced polymer composites (CFRP), aluminum (AL), and poly(methyl methacrylate) (PMMA, known as Plexiglas-PLEXI) and measured thermal heterogeneity in breast cancer screening test (DMR). We compared the appropriateness of these approaches to the similar state-of-the-art thermographic methods, such as PCT, CCIPCT, Sparse- PCT, NMF-gd, and NMF-nnls, among dissimilar thermal datasets. 
The results indicated the significant performance of semi-, convex-, and sparse- NMF in detecting defects (average accuracy levels of 68.6$\%$, 91.8$\%$, and 60.1$\%$, respectively) and preserved thermal heterogeneity to discriminate between symptomatic and healthy participants (accuracy of 74.1$\%$, 75.9$\%$, and 77.8$\%$, respectively). 
Sparse- NMF and sparse- PCT showed significant robustness against noise, which was tested by additive Gaussian noise from 0$\%$ to 20$\%$ than other thermographic techniques. 
Future works should substitute the manual selection of the basis from the low-rank matrix approximation with an automatic selection. In addition, an expansion of the validation set to a larger infrared imaging cohort can further confirm the strength and limitations of these approaches.

\section*{Acknowledgment}
This is the authors version of the article with doi: 10.1109/TIM.2020.3031129 in IEEE Transactions on Instrumentation and Measurement 2020.

\ifCLASSOPTIONcaptionsoff
  \newpage
\fi

\bibliographystyle{IEEEtran}
\bibliography{IEEE-NMF-bib.bib}

\appendices
\end{document}